\documentclass[twocolumn]{aa}
\usepackage{graphicx}
\usepackage{natbib}
\bibpunct{(}{)}{;}{a}{}{,}
\begin{document}
\title{A Mass Model for Estimating the Gamma Ray Background of the Burst and Transient Source Experiment }

\titlerunning{A Mass Model for Estimating the Background of BATSE}

\author{S.E. Shaw
\thanks{\emph{Present address:} INTEGRAL Science Data Centre, Chemin d'Ecogia 16, CH-1290 Versoix, Switzerland}
\and M.J. Westmore
\and A.J. Bird
\and A.J. Dean
\and C. Ferguson
\newline
\and J.J. Lockley
\thanks{\emph{Present address:} Oxford University Computing Centre, Wolfson Building, Parks Road, Oxford, OX1 3QD, UK}
\and D.R. Willis }

\offprints{S.E. Shaw, \email{simon.shaw@obs.unige.ch}}

\institute{Department of Physics and Astronomy, University of Southampton, SO17 1BJ, UK}

\date{Accepted 13 November 2002 by A\&A - this is the astro-ph version (compressed figs)}

\abstract{
Orbiting x-ray and gamma ray instruments are subject to large background count rates due to local particle fluxes in the space environment.  The ability of an instrument to make calibrated measurements of the flux from a source of interest is highly dependent on accurately determining the background level.  We present here a method of calculating the energy dependent background flux for any point in the complete data set recorded by the Burst and Transient Source Experiment (BATSE) in its nine year mission.  The BATSE Mass Model (BAMM) uses a Monte Carlo mass modelling approach to produce a data base of the gamma ray background which is then filtered to simulate the background count rate with a 2.048 second time resolution.  This method is able to reduce the variations in the background flux by a factor of 8 - 10, effectively `flat-fielding' the detector response.  With flat-fielded BATSE data it should be possible to use the Earth occultation technique to produce a hard x-ray all sky survey to the 1-2 mCrab sensitivity limit.  BAMM is also capable of estimating the contribution to the spectra measured from gamma ray sources due to the reprocessing of source photons in inactive material surrounding a gamma ray detector.  Possible applications of this aspect of the model in the area of Gamma Ray Burst spectral analysis are discussed.
   }

\maketitle

%

\section{Introduction}

Instruments in orbit around Earth encounter a hostile evironment of high
energy photons and cosmic ray particles.  In a Highly Elliptical Orbit a spacecraft is subjected to high fluxes of cosmic ray particles and gamma ray photons, which remain constant for long portions of the orbital period.  For Low Earth Orbits (LEO) the natural shielding from charged particles due to the local geo-magnetic field ({\em rigidity}, see section \ref{sec:geomag_effects}) is much higher, but background components are also highly variable on the orbital timescale which may be as short as an hour.  For applications in commercial satellites the deleterious effects of the background flux on electronics can be reduced by effective shielding.  For scientific instruments however the background flux is often the same energy as the photons of interest.  In order to make calibrated determinations of the flux and spectra of astrophysical sources it is important to accurately determine the background level.  This is especially true when observing the temporal characteristics of sources on timescales similar to that of the background variations.


\section{The Burst and Transient Source Experiment}

\begin{figure*}
\centering
\includegraphics[width=17cm]{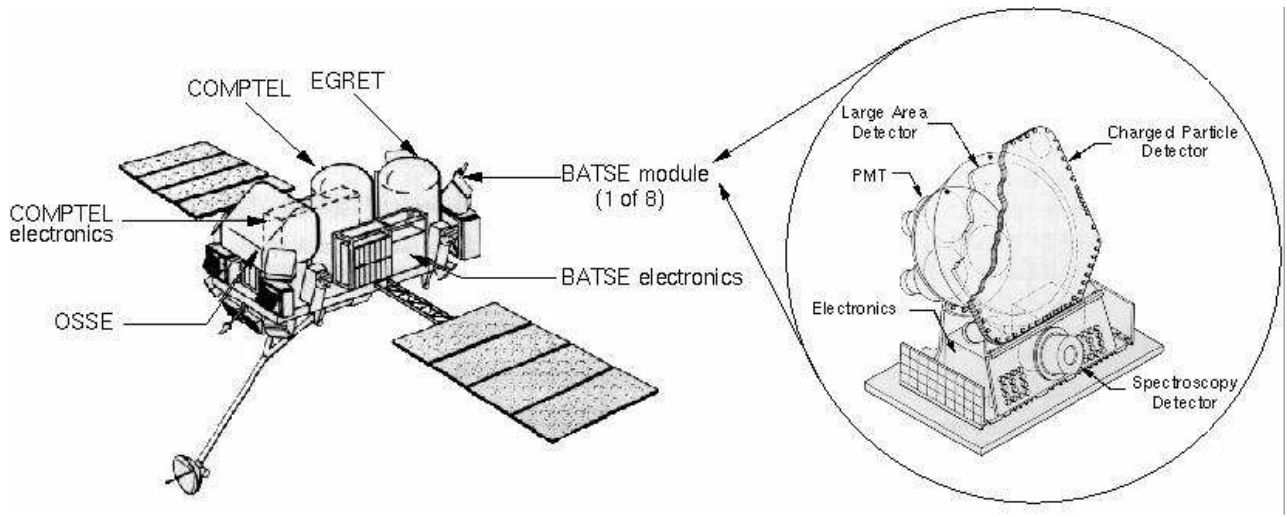}
\caption{The Compton Gamma Ray Observatory and one of the 8 BATSE detector modules (drawings courtesy of National Space Science and Technology Centre, Huntsville, AL, USA).}
\label{fig:cgro}
\end{figure*}

The Burst and Transient Source Experiment (BATSE) was an instrument
aboard the Compton Gamma Ray Observatory (CGRO) designed primarily to
detect and locate Gamma Ray Bursts (GRB).  It consisted of eight modules each situated on a corner of the CGRO cuboid, as shown in Fig.~\ref{fig:cgro}.  The primary instruments on each module were uncollimated
2025 cm$^{2}$ NaI(Tl) Large Area Detectors (LADs), all eight of which together had a total
field of view of $4\pi$ steradians and were sensitive from 20 keV - 2
MeV \citep{batse:fish89a}.  Data were recorded in several formats; the Discriminator Large Area (DISCLA) format contained 4 broad energy channels with data recorded every 1.024 seconds; the Continuous (CONT) data contained 16 energy channels with
a 2.048 second resolution.  Each BATSE module also included a Spectroscopic Detector (SPEC) situated just below each LAD and capable of measuring high resolution gamma ray spectra.  Lying immediately in front of and the same size as each LAD were the Charged Particle Detectors (CPD), a layer of plastic scintillator with the main purpose of acting as a veto shield for the LADs.  

BATSE monitored the whole sky continuously
for the $\sim$~9 year lifetime of CGRO and was an important
experiment in identifying the extragalactic origin of the Gamma Ray Bursts (GRB) \citep{batse:meegan92}.  

The 90 minute LEO of CGRO imposes
a variation on the background which fluctuates as the spacecraft moves
around the Earth.  The background is a complicated function of the
pointing of the detectors, the position of the spacecraft in the
Earth's magnetic field and its irradiation history.  With BATSE it is
vital to know the background and its fluctuations in order to make accurate measurements of gamma ray source fluxes, when using an observing technique such as the Earth Occultation Method \citep{harmon}.  Semi-empirical models are possible,
but they require assumptions about the physical processes involved or
calibration data taken from a detector after launch and may be subject
to large systematic errors.  A different approach is used here, based on the mass modelling technique.

\subsection{Previous BATSE Background Models}
The estimation of the background signal in gamma ray telescopes has historically been a question of informed guesswork.  Previous attempts to model the background count rate for BATSE detectors have been based largely on theoretical estimations or multi parameter fits to data (e.g. \citet{rubin96}, \citet{jpl2000}).  This approach can now be changed due to the reduced cost of aquiring high performance computers and the availability of sophisticated particle physics code, such as the GEANT package \citep{geant}.  In the mass modelling approach a geometrical computer model of a spacecraft is constructed including details of the instrument's mechanical structures and the chemical composition of its component parts.  A Monte Carlo numerical simulation can then be used to estimate the background produced when high energy particles interact in the detector model.  A detailed review of mass modelling techniques, including applications for BATSE and other instruments can be found in \citet{ssr2002}.

\section{BATSE Static Mass Model}

\begin{figure*}
\centering
\includegraphics[width=17cm]{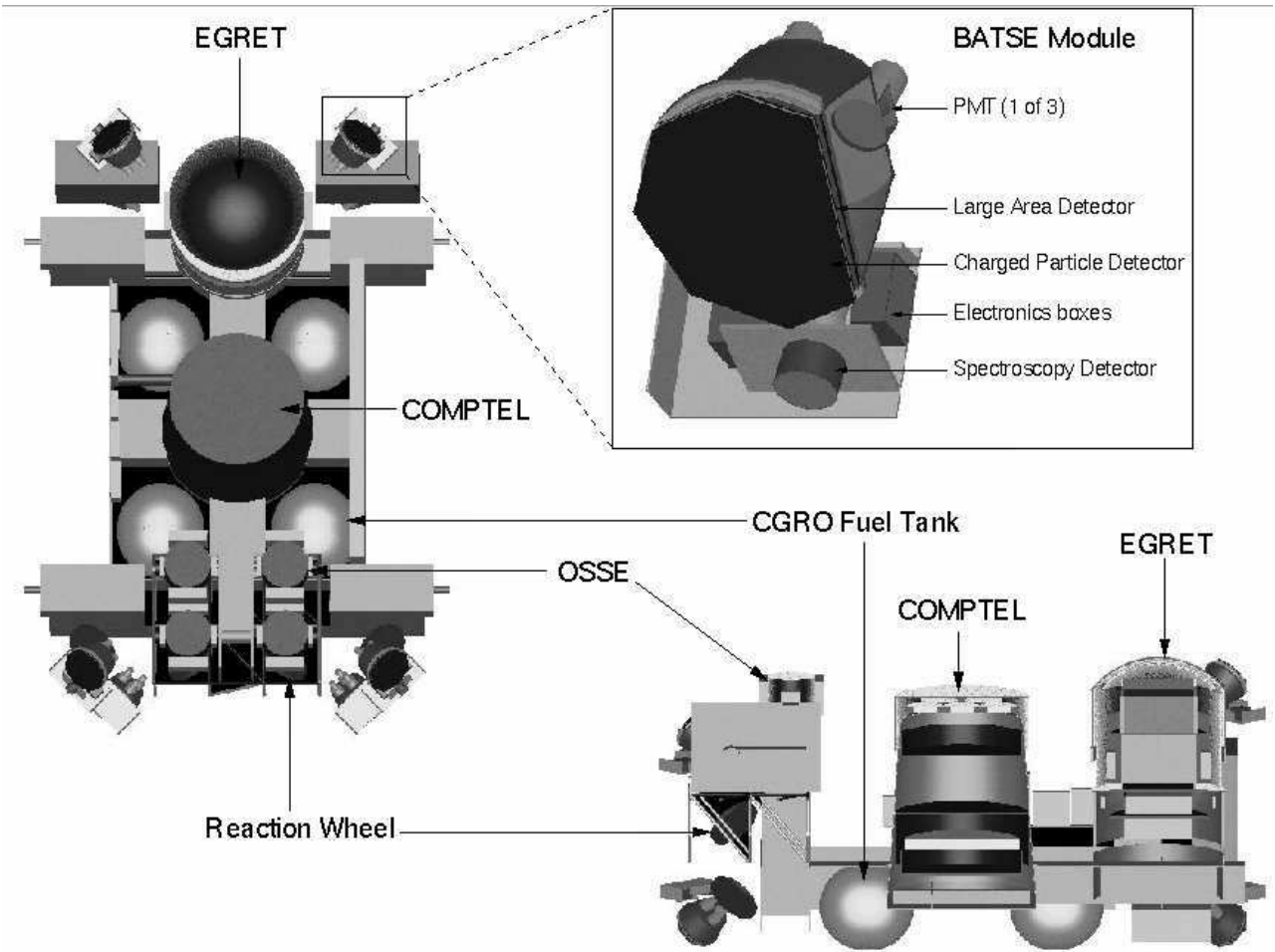}
\caption{BAMM contains detailed information about the BATSE modules (inset cut away image).  Also included in the model are the structure and chemical composition of the CGRO spacecraft and the major components of the OSSE, COMPTEL and EGRET instruments.  The lower image is a cut away along the long axis of CGRO and, to further optimise the view, some of the external volumes and surfaces of the model have been made transparent.}
\label{fig:cgromm}
\end{figure*}

The GEANT3 Monte Carlo particle interaction simulation software package has been used for several years in Southampton and in other groups to model astronomical gamma ray detectors, such as the INTEGRAL Mass Model (TIMM), described in \citet{timm:lei99}.  GEANT3 is a suite of software created originally for high energy particle physics experiments at CERN, in which a numerical model of a detector can be built and subjected to fluxes of particles \citep{geant}.  The suite of software includes detailed physical routines that use Monte Carlo techniques to simulate the trajectories, interactions and decays of sub-atomic particles within the model.  Components within the detector are modelled as geometrical volumes with detailed information about their chemical composition (to allow accurate interaction cross sections and isotopic decays to be modelled by the suite of physics routines).  The availability of computing resources and time limit the complexity of the model that can be created.  Compromises can be made, however, since the components of most interest are those active parts of a detector or telescope that would record information about the photons or particles in the actual instrument.  The probability of particle interaction depends approximately on the mass of the target volume and with each subsequent interaction or decay the spatial distribution of the resultant particles tends to isotropise.  Hence the level of detail required for components within the model scales roughly as the ratio of the component's mass to the square of the distance of that component from an active detector element.

The BATSE Mass Model (BAMM) has been constructed using GEANT to examine the effect of the space radiation environment on the detectors and has been adapted from earlier work within the group at Southampton \citep{deanrg2000}.  BAMM, shown in Fig.~\ref{fig:cgromm}, is constructed from $\sim 2000$ geometrical volumes representing the components in the entire CGRO spacecraft.  It was possible to obtain detailed information about the construction of BATSE, CGRO and the other instruments from engineering drawings. Each BATSE detector module consists of $\sim 100$ individual volumes whose size, shape and chemical composition were recreated as accurately as possible.  

Initially the model is assumed to be in deep space, away from any other object, and immersed in simulated isotropic particle fluxes.  This first stage of the modelling process records energy deposits from particles interacting in various volumes marked as `active' in the model.  Of primary interest for this work are the Na(Tl) scintillator crystals within the BATSE LADs, although particles interacting in the smaller SPEC detectors and the CPDs are also recorded.  The information recorded for any interaction in an active volume contains a history, including any previous scatterings, hadronic interactions etc., all the way back to the initial particle.  This information is used to determine the initial particle's trajectory and energetics for use later on in the background simulation process.  Three main components are considered in BAMM; the cosmic diffuse gamma rays (CD), prompt interactions of the cosmic rays with the spacecraft (CR) and the atmospheric albedo gamma rays (ATM). 

\subsection{Cosmic Diffuse Gamma Ray Spectrum}
Since there is no way of distinguishing a Cosmic Diffuse background gamma ray photon (CD) from a genuine source photon, measuring the CD background flux component is imperative in understanding the response of a detector.  The isotropic CD background is given in photons~cm$^{-2}$s$^{-1}$MeV$^{-1}$sr$^{-1}$ by the following spectra \citep{gehrels91};

\begin{equation}
\frac{dI_{\rm CD}}{dE} = \left\{ \begin{array}{lll}
	 0.54&E_{\gamma}^{-1.40} & \mbox{ for $10 < E_{\gamma} < 20$ keV}\\
	 0.0117&E_{\gamma}^{-2.38} & \mbox{ for $20 \leq E_{\gamma} < 100$ keV}\\
	 0.014&E_{\gamma}^{-2.30} & \mbox{ for $0.1 \leq E_{\gamma} < 10$ MeV}
	\end{array}
\right.	
\label{eqn:cd}
\end{equation}

BAMM is subjected to an isotropic flux of simulated gamma rays drawn randomly from the above spectrum. 

\subsection{Cosmic Ray Spectrum}
The Cosmic Ray induced background component (CR) is simulated from an incident flux of cosmic ray protons given by the following spectra in protons cm$^{-2}$s$^{-1}$MeV$^{-1}$ \citep{weber1974};

\begin{equation}
\frac{dI_{\rm CR}}{dE_{p}} = \left\{ \begin{array}{cl}
	 3.44\times 10^{2}E_{p}^{-2.04} & \mbox{ for $3.44 < E_{p} \leq 25.5$ GeV}\\
	 2.44\times 10^{5}E_{p}^{-2.69} & \mbox{ for $E_{p} > 25.5$ GeV}
	\end{array}
\right.
\label{eqn:cr}
\end{equation}

The CR spectrum is assumed to be isotropic and is valid throughout the solar cycle in the energy ranges of interest.  The CR component is the result of prompt interarctions in BAMM and includes gamma ray interaction products such as Compton scattered photons.

\subsection{Atmospheric Albedo Gamma Rays}

\begin{figure}
\resizebox{\hsize}{!}{\includegraphics{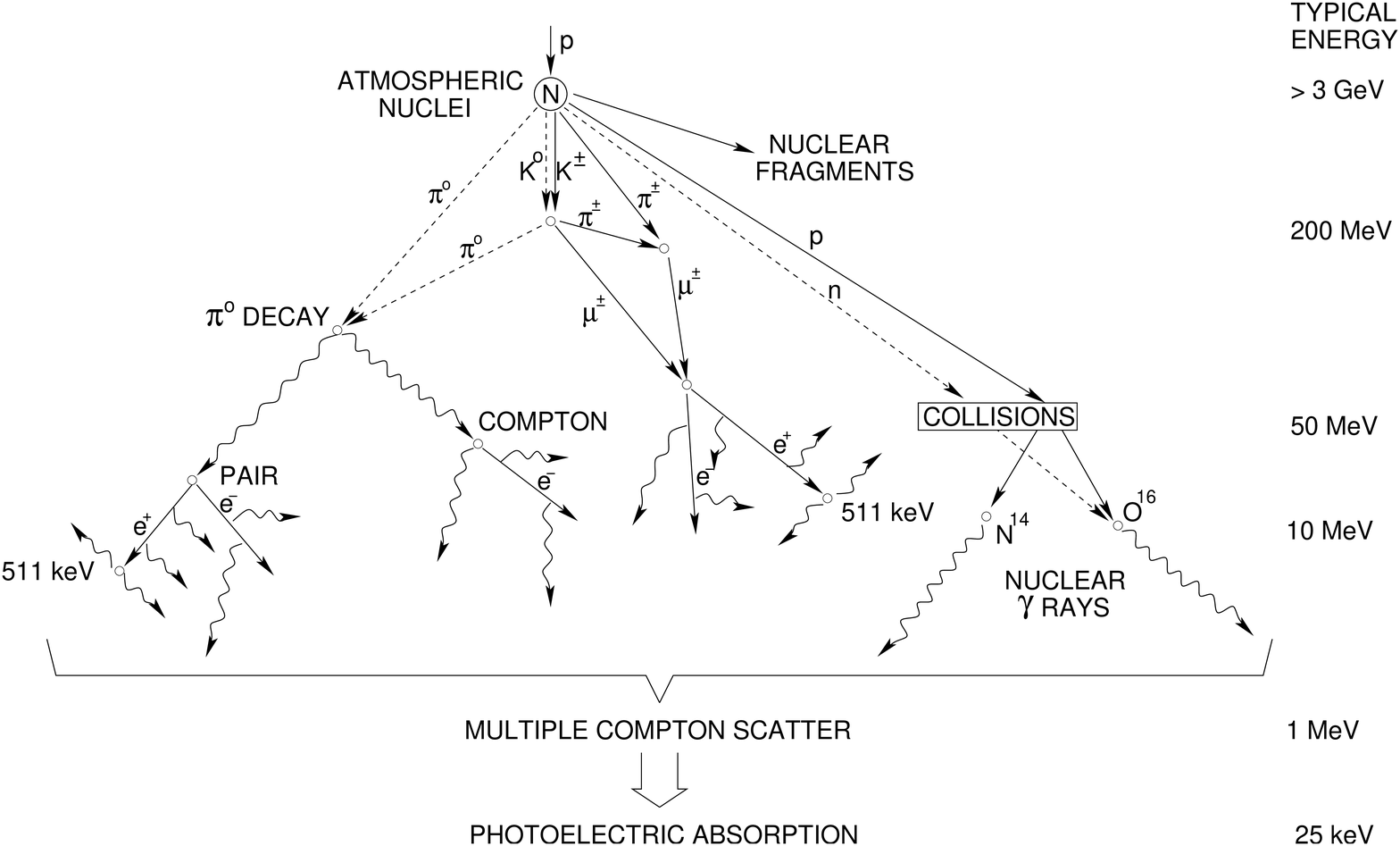}}
\caption{The mechanisms behind the production of the atmospheric albedo gamma ray flux.  Charged particles are indicated by solid lines whereas neutral particles are shown using dashes (from \citet{zombeck}).}
\label{fig:atmprod}
\end{figure}

\begin{figure}
\resizebox{\hsize}{!}{\includegraphics{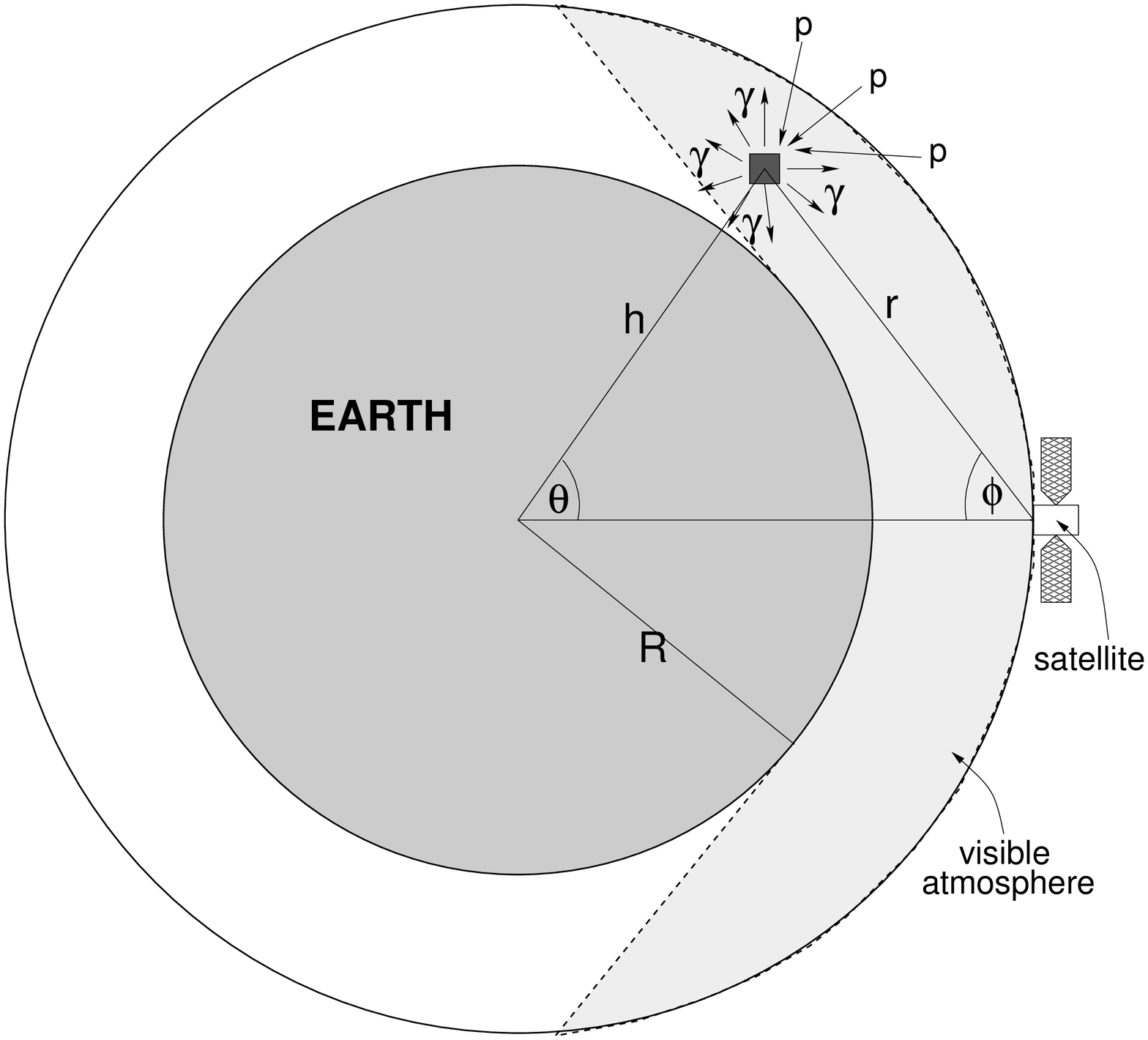}}
\caption{A schematic illustrating the terms used in the atmospheric albedo gamma ray emissivity function given in equation \ref{eqn:visatm}.}
\label{fig:atmearth}
\end{figure}

The Atmospheric Albedo flux of gamma rays is produced by cosmic ray interactions in Earth's atmosphere.  A schematic diagram of the process is shown in Fig.~\ref{fig:atmprod}. Pionisation of atmospheric nuclei by the incident cosmic ray flux leads to the production of muons, nuclear fragments, and other hadrons.  Gamma rays, above $\sim$ 50~MeV are produced directly by the decay of neutral $\pi^{\circ}$ particles (rest mass $\sim 140$ MeV).  The charged pion component also produces electrons, via the decay of charged muons, which radiate photons by bremsstrahlung.  The bulk of the gamma ray flux below 50~MeV is due to the bremsstrahlung photons and Compton scattering of higher energy photons. 
Since the photons are produced by secondary processes the flux emitted by a unit volume of atmosphere is spatially isotropic.
The energy spectrum of the atmospheric albedo gamma ray flux has been measured accurately, with minimum contamination from the cosmic diffuse gamma ray background, in balloon flights over Palestine, Texas, USA at an atmospheric depth of 2.5 g cm$^{2}$ (\citet{schon77}, \citet{schon80}).  The depth dependent gamma ray flux, measured in photons cm$^{-2}$s$^{-1}$keV$^{-1}$g$^{-1}$ at an atmospheric depth of $\chi$ g cm$^{2}$ is;

\begin{equation}
\frac{dI_{\rm ATM}}{dE_{\gamma}} = 0.535 \chi E^{-1.65}_{\gamma}.  
\label{eqn:atmspec}
\end{equation}

\begin{figure}
\resizebox{\hsize}{!}{\includegraphics{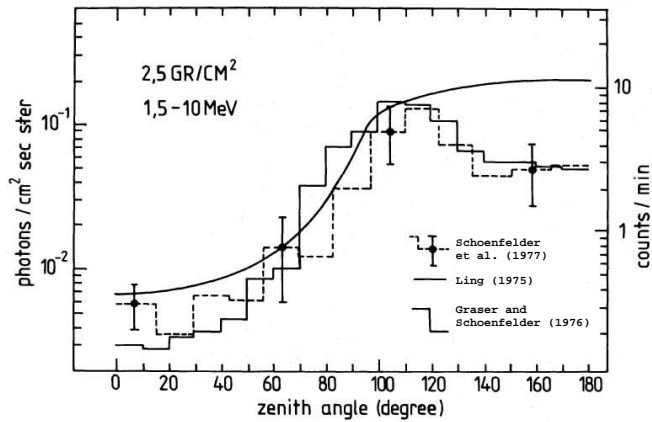}}
\caption{The zenith angle distribution of the 1.5 - 10 MeV atmospheric albedo gamma ray flux at an atmospheric depth of 2.5 g cm$^{-2}$ (42 km altitude) and a local rigidity cutoff of 4.7~GV, taken from \citet{schon77}.  An angle of $0^{\circ}$ in the diagram indicates downward moving photons (towards the Earth) where as upward moving photons have an angle of $180^{\circ}$.  The flux measurements of \citet{schon77} are compared with those of \citet{ling75} and \citet{graser77}.}
\label{fig:atmzen}
\end{figure}

 Unlike the isotropic distribution of particles used when BAMM simulates the CD and CR components the ATM gamma rays are, in effect, emitted by the Earth's disk.  The total gamma ray emission is not isotropic over all zenith angles.  The zenith angle distribution (measured by \citet{schon77}) is shown in Fig.~\ref{fig:atmzen}.  The peak at $\sim 120^{\circ}$ is due to photons that were emitted in a preferentially horizontal direction ($90^{\circ}$) but experienced Compton scattering between atmospheric element and detector.  

The integrated emissivity model of \citet{dean89} is used here to calculate the integrated atmospheric emission of the albedo flux over the volume of atmosphere visible to the spacecraft, $V$.  The line of sight dependent absorption between the space craft and an atmospheric element for photons of energy $E_{\gamma}$, characterised by the photon attenuation coefficient $\mu(E_{\gamma})$, is given by 

\begin{equation}
M = e^{\left[-\mu(E_{\gamma})\int_{r}\rho(r)dr\right]}.
\label{eqn:m}
\end{equation}

Hence the atmospheric gamma ray flux incident on a satellite is given by,

\begin{equation}
\frac{dI_{\rm ATM}}{dE_{\gamma}} = \int_{V}\frac{2\pi h \sin\theta \rho(h) \epsilon[E_{\gamma},h]h^{2}\cos\phi}{4\pi r^{2}}\cdot M \cdot d\theta dh .
\label{eqn:visatm}
\end{equation}

The atmospheric air density, $\rho$, is described by two functions in $r$ and $h$, where Fig.~\ref{fig:atmearth} shows the definitions of $h, r, \theta$ and $\phi$.  The depth dependent mass emissivity function, $\epsilon[E_{\gamma},h]$ photons s$^{-1}$keV$^{-1}$g$^{-1}$, is based on the spectrum of equation \ref{eqn:atmspec} and takes the form;

\begin{equation}
\epsilon(E_{\gamma},\chi) = 0.75\chi^{0.51}E_{\gamma}^{-(1.65+2.5610^{-2}\chi^{0.50})}.
\label{eqn:massemiss}
\end{equation}

The emissivity function depends on the incident cosmic ray flux, which is shielded at low energies by the Earth's magnetic field.  The measurements given above were made over Palestine at a local rigidity of 4.7 GV.  Treatement of the ATM component as the local value of the minimum geo-magnetic rigidity cut off changes in orbit is discussed further in section \ref{sec:ridgatm}.

 Changes in altitude, $h$, affect the solid angle subtended by the atmospheric disk and hence the strength of the albedo flux.  The orientation of the detector with respect to the geo-centre also reduces the photon flux as the atmosphere leaves the field of view.  It would not be computationally possible to model every possible combination of altitude and pointing direction for the whole CGRO model.  Instead a data base of responses from a single BATSE module is created at discrete altitudes (every 50 km from 350 - 850 km) and attitudes (every 5$^{\circ}$ from 0 to 180$^{\circ}$).  In the CD and CR component modelling we are interested not only in the direct effects of background photons interacting in the detector crystals, but also in the secondary radiation produced by high energy interactions in other parts of CGRO finding its way to the detector.  However, the photons of interest in the ATM component are largely detected by direct interaction in the detectors themselves, the chance of a 2 MeV photon interacting in CGRO and producing a detectable photon in BATSE above 20 keV is small.  Modelling the BATSE detector alone has the obvious advantage of a large reduction in computing time, since the model is being run a total of over 400 times. 

\section{Spatially and Temporally Dynamic Mass Model}
The BAMM results are `static', they contain a data base of gamma rays and cosmic ray particles arriving from all directions and a data base of atmospheric albedo gamma rays for a grid of LAD pointing angles and orbital altitude.  The second stage of the process is to remove those particles from the data base that are unphysical for particular CGRO orientations and positions.  The major effect on the background is the motion of CGRO around Earth modulating the component flux.  The CR and CD components reduce as the Earth enters the LAD field of view and blocks the incident flux.  The ATM component can be thought of as the emission of gamma rays from a spherical shell and so the component flux increases as the atmospheric column density increases, i.e. as the limbs of the Earth enter the field of view.  The second major effect is that of the local value of the rigidity of the Earth's magnetic field, which shields charged particles from CGRO and hence affects the CR and ATM components.  Although the ATM component is effectively dependent on the CR flux, it is highly unlikely that a cosmic ray that interacts in the atmosphere and produces albedo photons will also go on to reach CGRO.  Hence the accuracy of the model is not reduced by the decoupling of these two components.

\subsection{Attenuation of Components due to Obscuration by the Earth} 
\label{sec:earthblocking}
The CD and CR results from the static mass models contain a set of energy deposits in the BATSE LADs and records the momentum of the initial particle, which was drawn from a spatially isotropic distribution.  The low Earth Orbit of CGRO was circular with an altitude of 400 - 500 km.  Seen from this distance the Earth's disk subtended an angular radius of $\sim 70^{\circ}$ at the spacecraft, hence $\sim$~30\% of the sky was blocked by the Earth at any one time.  Consequently many of the energy deposits recorded in the static mass model stage come from particles arriving from non-physical directions and must be filtered from the data set.

Since one of CGRO's tasks was to make measurements of pulsars, where precise positioning is vital for barycentering pulse arrival times, the location of the spacecraft is accurately recorded in the telemetry data.  This, along with the orientation of the spacecraft, allows the position of the Earth as seen from each of the eight LADs to be calculated.  Hence, given a particular part of the CGRO mission lifetime with its position information, the CD and CR components can be calculated by finding the Earth's position and removing events due to Earth blockage.  This leaves behind background components modulated for the effect of the Earth obscuring events from reaching the LADs.

\subsection{Effects of the Geo-Magnetic Field on Background Components}
\label{sec:geomag_effects}
Both the CR and the ATM components are derived from the incident isotropic proton cosmic ray flux. They are therefore both modulated,
 not only by obscuration by the Earth, but also by the shielding of charged particles by the Earth's magnetic field.  A cosmic ray's ability to penetrate the Earth's magnetic field is characterised by its magnetic rigidity, $R$, a measure of the particle's momentum per unit charge.  An ion with atomic mass number $A$, charge number $Z$ and kinetic energy $E$ (in units of GeV/nucleon) has rigidity

\begin{equation}
R = \frac{A}{Z} \surd(E^{2} + 2m_{o}E),
\label{eqn:rigid}
\end{equation}

where $R$ is measured in GV and $m_{o} = 0.9315$ GeV/$c^{2}$ is the atomic mass unit.  A particle with $R$ less than the critical value, $R_{c}$, required by the local strength of the magnetic field will be deflected, it does not lose energy since the magnetic field acts in a direction perpendicular to the particle's momentum.  This effect shields the Earth from cosmic ray particles and, very simply, cuts out those particles with low rigidity.  The rigidity required by particles to reach CGRO in orbit varies as the spacecraft moves through the geo-magnetic field and also depends on the orbital altitude.  A simple equation can be used to estimate the local rigidity during a spacecraft's orbit based on the McIllwain $L$ parameter, $R_{c} = 15.6/L^{2}$ \citep{gehrels91}.  The $L$ parameter is a coordinate describing the effective orbital position, as a fraction of the Earth's equatorial radius,  in an ideal dipole field that has the same local magnetic field strength as the measurement.  For the circular LEO of CGRO at an altitude of $\sim$ 500 km, $L$ varies from $\sim 1$ above the equator to $\sim 1.5$ at the highest orbital latitude of $\pm 28^{\circ}$.  This gives the minimum rigidity required by a proton as $12 \geq R_{c} \geq 7$ GV, corresponding to an energy range of $13 \leq E \leq 8$ GeV/nucleon.

 Since the CR component is
 modelled directly from the initial cosmic ray input spectrum, the BAMM
 results can be filtered according to the local minimum rigidity. The
 ATM gamma ray component is not modelled directly from the incident cosmic
 rays and so a semi-empirical approach must be applied.  

\subsubsection{Rigidity Effects on the Cosmic Ray Component}
The shielding effect of
 the magnetic field can be approximated by an abrupt cut-off in the
 incident proton cosmic ray spectrum, i.e. the number of particles, $N_{B}$, that would be detected in a local field with strength $B$ is given by,
\begin{equation}
N_{B} = \int^{\infty}_{E_{c}}I_{\rm CR}  dE,
\label{eqn:cutoff}
\end{equation}

where $E_{c}$ is the minimum kinetic energy required by a charged particle to penetrate the magnetic field and $I_{CR}$ is the incident cosmic ray spectrum.

In order to apply a magnetic
 shielding algorithm, the minimum local rigidity must be known. In
 this case a semi-empirical measure is made. The BATSE detector modules were manufactured with plastic scintillator charged particle detectors (CPD) in front of each NaI LAD to act as a veto counter (recall Figs.~\ref{fig:cgro} and \ref{fig:cgromm}).  The CPD rates were recorded by BATSE on the same 2.048 second time resolution as the LAD counts, positional information etc.  Likewise, BAMM contains models of the CPD detectors and records particles that lead to an energy deposit within them in the same way that they are for the LADs.  The BAMM CPD counts are filtered to remove those events that would have been prevented from reaching the instrument by the Earth as described in section \ref{sec:earthblocking}.  Assuming that BAMM is a reasonable estimate of the CPD angular response, the number of charged particles remaining is that that would be detected by the CPDs in the absence of any magnetic field.  Using equation~\ref{eqn:cutoff} an approximation for the ratio of CPD counts observed by BATSE, $N_{\rm CPDobs}$, to those predicted by the model, $N_{\rm CPDmod}$, can be made;
\begin{equation}
\frac{N_{\rm CPDobs}}{N_{\rm CPDmod}} = \frac{\int^{\infty}_{E_{c}} I_{\rm CR} dE}{\int^{\infty}_{0}I_{\rm CR} dE}
\label{eqn:crcutoff}
\end{equation}

Solving the integral formed by substituting the cosmic ray spectrum and boundaries given by equation~\ref{eqn:cr} into equation~\ref{eqn:crcutoff} gives an expression for $N_{\rm CPDobs}/N_{\rm CPDmod}$ in terms of $E_{c}$.  This can be rearranged to give an expression for $E_{c}$;


\begin{equation}
E_{c} = \left[\frac{0.0601\left(\frac{N_{CPD obs}}{N_{CPD mod}}\right) - 3.743\times10^{-3}}{329.89}\right]^{\frac{1}{1.0421}}.
\end{equation}

Equation \ref{eqn:rigid} can be used to determine the value of $R_{c}$ from $E_{c}$.  At each 2.048s time interval, for which the position information for
CGRO is known, the value of $E_{c}$ is calculated and compared against the incident particle energy for each energy deposit predicted by BAMM. If the incident particle is determined to have energy less
than this minimum value it can never have reached a position to cause
an energy deposition and it is removed from the data base.

\subsubsection{Rigidity Effects on the Atmospheric Component}
\label{sec:ridgatm}
Background components that are dependent on the cosmic ray flux but
are not modelled from individual incident cosmic rays, such as the ATM
gamma ray component, must be treated differently.

An empirical rigidity correction factor, $M(R_{c})$, based on a number of
satellite observations, e.g. \citet{imhof76}, may be defined as, 

\begin{equation}
M(R) = \int^{\infty}_{E_{c}}\frac{dI(E_{p})}{dE_{p}}E_{p}^{0.7}dE_{p},
\label{eqn:rigidcorr}
\end{equation}

where $dI(E_{p})/dE_{p}$ is the cosmic ray intensity at proton energy $E_{p}$.  Numerous measurements have been made of $R_{c}$ above Palestine, Texas, of 4.7 GV.  A correction factor, $f(R) = M(R_{c})/M(R_{c}=4.7)$, is defined that can be used to scale the Palestine rigidity measurements to any value of $R_{c}$.  By solving the integral formed by substituting equation \ref{eqn:rigidcorr} with the cosmic ray spectrum and boundaries of equation \ref{eqn:cr}, the rigidity correction factor is,

\begin{equation} 
f(R) = 28.462 E_{c}^{-0.342} - 0.579.
\end{equation}
 
\subsection{Other Components}
The background components mentioned above are modulated on a psuedo periodic time scale, governed by the orbital motion of CGRO around the Earth, and are reasonably simple to predict.  Other particle processes exist that can produce a detectable background component with weak dependence on the orbital period.  These background contributions may become important for specific spacecraft positions and/or orientations and as such are harder to predict than the CD, CR and ATM components.  Two effects, due to the South Atlantic Anomaly (SAA) and atmospheric electron bremmstrahlung are considered here:

The very high fluxes of energetic protons found within the SAA ($> 1000$ protons~cm$^{-2}$s$^{-1}$ above 30~MeV) activate isotopes within the material of a spacecraft passing through it \citep[see for example][]{roederer70}.  These isotopes subsequently decay and produce a delayed count rate in gamma ray detectors that can last for tens of minutes after the initial pass.  The charged particles are also capable of severely damaging electrical equipment passing through them.  A typical solution, used with BATSE and CGRO, is to avoid damage by shutting down instruments during passage through the SAA (about 5-6 times during a typical 24 hour period).  This makes the estimation of the decay of the spectral count rates on emergence from the SAA unreliable.  The problem is further compounded since much of the incident particle flux has low penetrating power in the material of the spacecraft.  Therefore anisotropy of the trapped particle momentum distribution with respect to the CGRO orientation vector makes exact determination of the induced radioactivity, and hence the eventual background contribution, extremely difficult to model.

The Earth's magnetic field is also responsible for a small background component caused by aurora-like emissions of x-ray photons produced by electron bremsstrahlung above the North and South poles \citep{aschwandena,aschwandenb}.  For BATSE these emissions may be observed in the lowest energy channels by individual detectors during periods of the orbit when they are pointing towards the poles. The effect manifests itself as a short `blip' -- a jump of a few tens of counts that lasts $\sim$~5 minutes and appears two or three times a day in detectors alligned in favourable directions.  Since these events are short and infrequent they are not considered a serious problem.  Modelling the electron bremsstrahlung component would also be difficult since the strength of the effect is dependent on solar activity amongst other things.

A detailed study of these two effects has shown that a loss of only $\sim$ 15\% of usable data are achieved by not considering parts of the data that are severely affected by these two components \citep{kt,gb}.  Although it is in principle possible to use the mass modelling technique to produce these extra components it is not considered an efficient use of time for the present.

\subsection{Renormalization of Background Components}
The mass modelling technique outlined above produces three components, which when added together, gives a close approximation of the actual gamma ray background.  The accuracy of the technique is limited by how well the input spectra for the CD, CR and ATM components have been measured.  Due to uncertainty in the shape of the CD, CR and ATM input spectra the results of the mass modelling cannot exactly reproduce the actual count rates seen in the BATSE detectors.  Although the time profile of the counts seen in each detector is reproduced accurately the absolute normalization of each component is not.  Consequently it is necessary to rescale the three components.

A short section ($\sim$~5 orbits) for each detector and CONT energy channel in a time period of interest is chosen such that it is free from contamination due to SAA passages, known flaring gamma ray sources and large data gaps.  This section is then corrected by subtracting a contribution from all known gamma ray sources above 200~mCrab brightness.  A model of the form
\begin{equation}
{\rm C}(t) = \alpha \cdot{\rm CD}(t) + \beta \cdot{\rm CR}(t) + \gamma \cdot{\rm ATM}(t)
\label{eqn:renorm}
\end{equation} 

is fitted to the short section of corrected data using a least squares method.  The rescaling returns the global amplitudes, $\alpha, \beta, \gamma$,  of the CD, CR and ATM components respectively which are used to return the total background for the day, detector and channel in question.  It is important to highlight that the fitting routine in no way alters the predicted shape of the count rate profile of the three components returned by the mass model.  

\section{BATSE Dynamic Mass Model Results and Applications}

\begin{figure*}
\vspace{10mm}
\centering
\includegraphics[width=17cm]{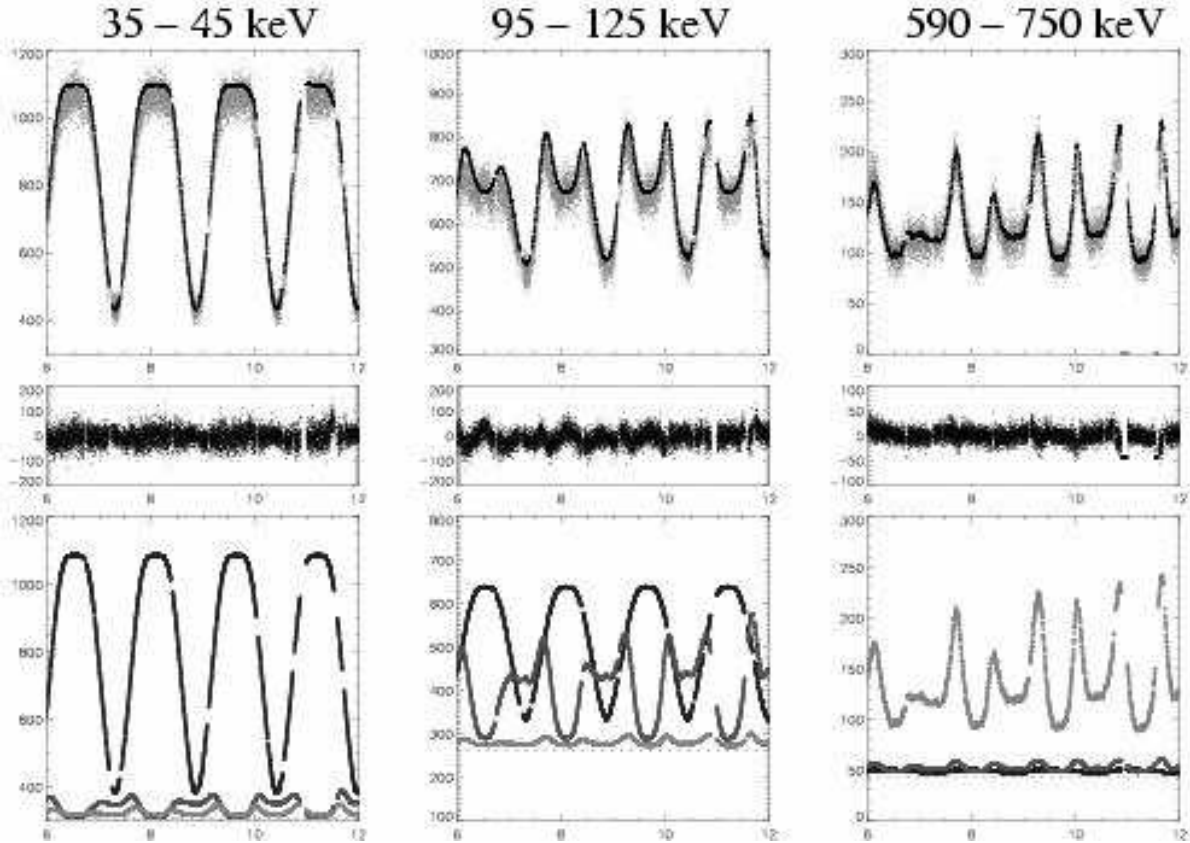}
\vspace{10mm}
\caption{BATSE LAD 5 dynamic background model for three different energy 
channels in a twelve hour section of the CGRO mission.  The top row of
figures shows the excellent agreement of the dynamically filtered BAMM (dark
line) with the BATSE CONT data (light dots).  The central
figures show the residuals of the above plots and highlight the large
reduction in the BATSE background fluctuation that can be achieved by
subtracting the model.  The bottom diagrams show the CD
(black), ATM (mid grey) and CR (light grey) components of the total
background model and a DC offset due to unresolved background
gamma ray sources (dotted line).  The dependence of the three components on
energy can clearly be seen.} \label{fig:bgcomp}
\end{figure*}

Figure~\ref{fig:bgcomp} shows the excellent
agreement between the dynamic BAMM data and the CONT count rate profile in the
BATSE LADs for three energy channels. The data are flat-fielded simply
by subtracting the modelled background, which reduces the fluctuation
of the background counts by up to an order of magnitude.  The relative importance of the three components as a function of energy is clear;  at low energies the CD component dominates the total background; at the high energy end the CR component is most important; the ATM component is at its strongest at intermediate energies.  The figure also shows the importance of modelling the detector attitude correctly.  The CD component shows a large modulation, indicating that the detector was pointing almost directly in the plane of the CGRO orbit.  As the pointing angle moves away from the orbital plane less of the Earth comes into the detector field of view and so less of the component will be blocked out.  In this case the component will have a similar maximum strength but much reduced variation.  Two applications of the flat-fielded data are discussed below.

\subsection{Gamma Ray All Sky Survey}
\label{sec:ass}

\begin{figure}
\resizebox{\hsize}{!}{\includegraphics{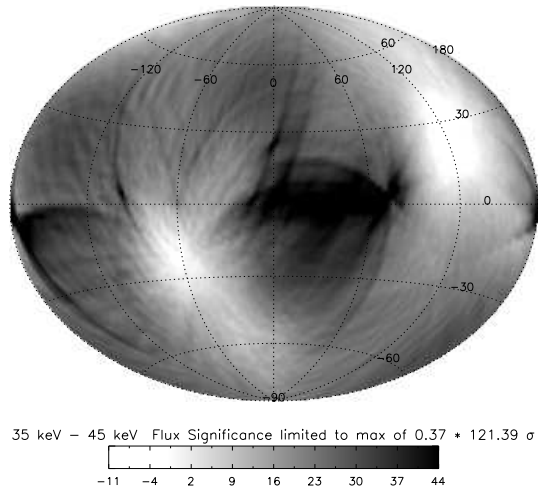}}
\caption{Maximum likelihood all sky image, displayed in Galactic Coordinates and showing the effect of applying the LIMBO technique to raw data.  The image is  made from CONT energy channel 2 ($\sim$~35 - 45 keV) for the period TJD
11400 - 11450 and the peak flux significance indicated is at a position consistent with the Crab Nebula, $(l,b) = (-175.4,-5.8)$.  Note the large systematic background, present at a level corresponding to $\sim$~25\% of the Crab flux.}
\label{fig:lumpy}
\end{figure}

\begin{figure}
\resizebox{\hsize}{!}{\includegraphics{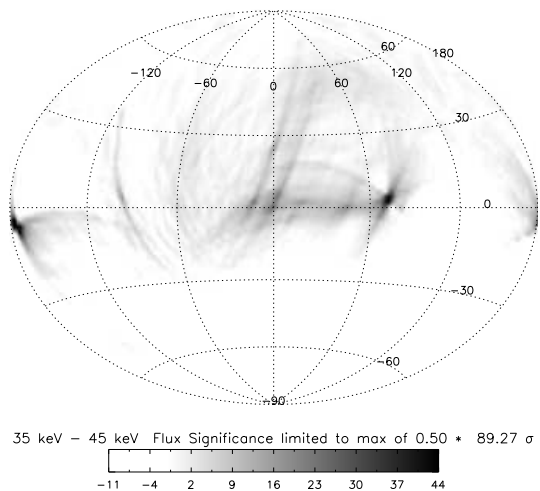}} 
\caption{Maximum likelihood all sky image, as Fig.~\ref{fig:lumpy}, but applied to data flat-fielded with dynamic BAMM data.  The same greyscale has been used in both figures.  The systematic background apparent in Fig.~\ref{fig:lumpy} has been reduced significantly.}
\label{fig:flat}
\end{figure}

The major driving force behind developing the dynamic mass model for BATSE was to enable the systematic errors in long ($\sim$ hundreds of days) sections to be reduced such that it could then be used for continual observations of objects with improved sensitivity.  The most important application of this is in the production of a hard x-ray all sky survey using the Earth occultation technique \citep{batse:fish89b,harmon}.  A maximum likelihood method is used to produce all sky images in the gamma ray band, the first time this has been achieved since the HEAO missions of the late seventies, described in \citet{HEAO1A4}.  The Likelihood Imaging Method for BATSE Occultation data (LIMBO) are discussed in full in \citet{shaw:alicante,mlmpaper}.

The flat-fielded data was shown to have much reduced background
variation in figure~\ref{fig:bgcomp} and thus should produce much
cleaner images.  Shown in Figs.~\ref{fig:lumpy} and \ref{fig:flat} are the results of applying LIMBO to the raw and flat-fielded BATSE count rates respectively. The maps were created for one CONT energy channel ($\sim$~35 - 45 keV) and summed
for the 50 day period from TJD 11400 - 11450.  

A severe and varying systematic background is present across the map made from the raw data.  Near the Crab Nebula position it is at a level equivalent to $\sim 30 \sigma$ or $\sim
250$~mCrab. In contrast the flat-fielded map shown in Fig.~\ref{fig:flat} has a drastically reduced systematic level that, at worst, is $\sim$~ 50 mCrab.  By simple extrapolation of the significance of the measurement of the Crab Nebula flux (but including any systematic offset) it is possible to estimate the statistical 3~$\sigma$ flux limits for the maps: $\sim$~25~mCrab and $\sim$~30~mCrab respectively for the raw and flat-fielded images.  The systematic error is clearly the dominant factor in determining the sensitivity of the imaging process.  When the entire BATSE data base is considered we expect to produce an all sky survey map with a sensitivity of $\sim$~1-2~mCrab.

The faint features stemming from bright point sources in the
flat-fielded map are due to the imaging process itself.  Individual
point sources are created by the superposition of many images of the
Earth's limb (radius $\sim 70^{\circ}$).  Therefore it is possible for a
bright source to have an effect many tens of degrees away from its
position on the sky.  The `point spread function' of a single source
is a complicated function of many parameters, but it is calculable.
Further reduction of the systematic background in the images, by producing methods of image cleaning the occultation maps
and by improving the background modelling, remains ongoing work \citep{mlmpaper,mattsthesis}).

\subsection{Gamma Ray Burst Studies}
In BAMM the accurate geometrical and chemical representation of the BATSE modules and their immediate neighbourhood (i.e. the entire CGRO spacecraft) may have further useful applications in the analysis of Gamma Ray Bursts (GRB).  The  GRBs detected by BATSE emit up to $10^{54}$ ergs above $\sim$~30 keV (postulated for isotropic emission) over a range of times as short as a few hundredths up to $\sim$~1000 seconds \citep{fishnchipsgrb,meszaros}.  Despite the higher signal to background ratio seen in GRB detections than in other types of gamma ray sources a background model is still important since the fluxes in GRB lightcurves can vary greatly.  It is also tantalising to be able to investigate the most distant GRBs that are detected at the threshold of sensitivity.  

For the most part the short length of the burst is such that the background does not vary appreciably and can be estimated by fitting simple empirical models to the data (e.g. \citet{2000ApJS..126...19P}).  However for the fraction of GRBs that last for more than a few 100 seconds these models start to run into difficulties as the background starts to vary.  In addition gaps in the BATSE telemetry become more likely the longer the time being considered and this further compounds problems with the fit.  

On timescales of many hours to perhaps a few weeks the measurement of GRB tails has become one of great interest \citep{1999ApJ...524L..47G}.  The current method of estimating the background for these events has involved interpolation of BATSE data from orbits before and after the orbit of interest \citep{2002ApJ...567.1028C}.  This method has shown interesting results, but is subject to systematic error in determining the count rate due to the precession of CGROs orbit and can be affected by change in the spacecraft conditions such as a reorientation to a new pointing.  

\begin{figure}
\resizebox{\hsize}{!}{\includegraphics{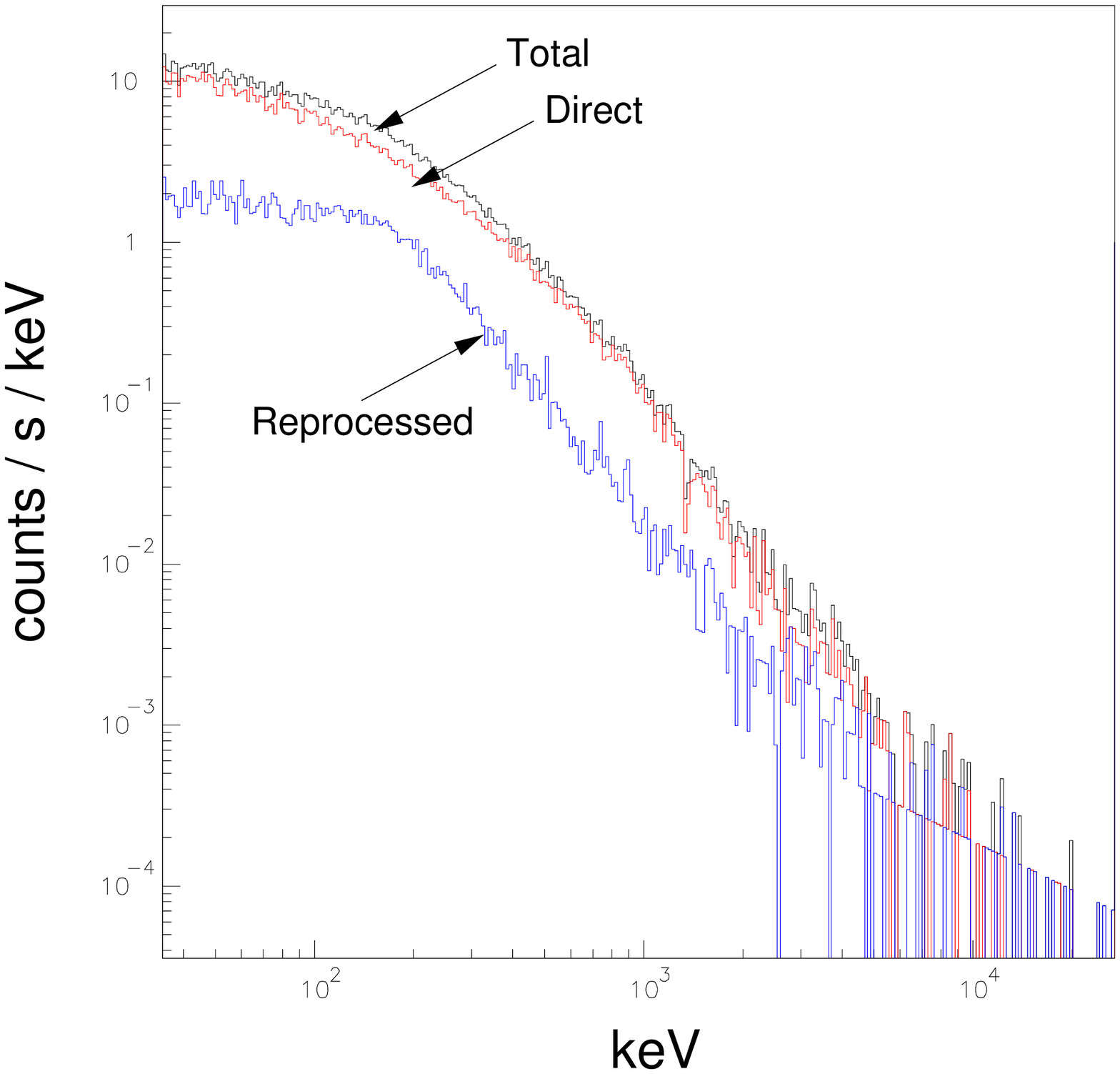}}
\caption{The modelled response of one of the BATSE SPEC spectrometer detectors to a GRB.  In this simulation the input spectrum was based on the spectrum measured for GRB91053 (Parameters provided by \citet{briggs} and using the model of \citet{bandeq}: $A = 0.118$ photons cm$^{-2}$s$^{-1}$keV$^{-1}$, $E_{o} = 753$ keV, $\alpha = -0.907$, $\beta = -2.72$).The figure suggests that 10-30\% of the total detected flux is not detected directly from the gamma ray burst but is in fact due to secondary photons emitted when gamma ray photons interact in the material of the space craft.}
\label{fig:grb}
\end{figure}

Spectral studies of GRBs is also an area of high scientific interest and a great deal of effort has been put in to searching for lines and other features in BATSE GRB spectra (e.g \citet{1998hgrb.symp..299B}, \citet{grb1} and others).  Knowledge of the spectral response of the detectors is highly important in deconvolving GRB spectra.  However, this is difficult to estimate due to the partial deposition of energy by high energy photons in the detector material \citep{1996hgrb.symp..133B} and the detection of secondary radiation created by Compton scattering of the source photons in the neighbourhood of the detectors, such as the spacecraft material or Earth's atmosphere \citep{1996hgrb.symp..182H}.  \citet{batseresponsematrix} have used a Monte-Carlo model to estimate the Detector Response Matrices (DRM) of the LAD and SPEC detectors.  This model assumed azimuthal symmetry of the detector response, which has limited the DRMs use with SPEC data to events occuring at no more than 60$^{\circ}$ from the detector axis.  The representation of the surrounding CGRO instrument was also fairly crude and the model is not able to account for the effect in all eight of the detector modules together and in situ on CGRO for different orientations and in different magnetic field conditions.

BAMM may be a very useful tool in future GRB analysis.  With a rigourous treatment of the effects on the spectrum due to the surrounding material, it will allow the measurement of the gamma ray background in all eight BATSE detectors for any position and orientation in the orbit history on any timescale down to the maximum time resolution of the dataset.  As an example Fig.~\ref{fig:grb} shows the spectrum calculated from one of the SPEC detectors by BAMM when presented with a simulated flux of gamma ray photons from a GRB.  A significant fraction ($\sim$ 10 - 30 \%) of the total flux is due to secondary photons and would lead to serious systematic errors when calculating the spectrum of the GRB.  The high level of spatial, mechanical and chemical information within BAMM together with the particle interaction simulation capabilities of GEANT3 also allows particles to be tracked in detail through the detector.  For example it is possible for BAMM to identify the high energy particles that deposit a small amount of energy, leave the detector and are then detected again after further scattering interactions elsewhere in the instrument.  Work continues to investigate to what level BAMM can be used and how the mass modelling technique can be adapted to other missions, such as the BAT GRB monitor onboard the SWIFT mission \citep{davesthesis,swift3}.   

\begin{acknowledgements}
The authors would like to thank G.J. Fishman, C.A. Wilson, M.S. Briggs and others at the National Space Science and Technology Center, Huntsville for their considerable assistance in providing data, gamma ray source measurements and GRB spectra and for their helpful comments during the preparation of this work.  The efforts of R. Gurriaran and F. Lei in the early stages of this project are recognised and greatly appreciated.
\end{acknowledgements}

\bibliographystyle{aa}
\bibliography{ms3051}

\end{document}